\documentclass[prd,a4paper,nofootinbib,
showpacs, 
]{revtex4}
\usepackage{graphicx}
\usepackage{amsfonts}
\usepackage{amssymb}
\def\eq#1{{eq.~(\ref{#1})}}
\def\eqs#1#2{{eqs.~(\ref{#1})--(\ref{#2})}}

\def\hbar{\hspace{0pt}\raisebox{1pt}{$-$} \hspace{-7pt} h}

\def\5{\overline 5}

\newcommand{\be}{\begin{equation}}
\newcommand{\ee}{\end{equation}}
\newcommand{\bea}{\begin{eqnarray}}
\newcommand{\eea}{\end{eqnarray}}
\newcommand{\nn}{\nonumber}

\begin{document}
\title[]{Telling the spin of the ``Higgs boson'' at the LHC\\
{\small Spin determination for resonances in $ZZ$, $W^+W^-$ and $t\bar{t}$ decay channels}
}
\date{\today
}
\author{U. De Sanctis$^{\dag *}$}
\author{M. Fabbrichesi$^{\ddag}$}
\author{A. Tonero$^{\dag\ddag}$}
\affiliation{$^{\ddag}$INFN, Sezione di Trieste}
\affiliation{$^{*}$INFN, Gruppo collegato di Udine} 
\affiliation{$^{\dag}$SISSA, via Bonomea 265, 34136 Trieste, Italy}

\begin{abstract}
\noindent   We assume that  the Higgs boson or a possible resonance---playing its role in strongly interacting models of electroweak symmetry breaking---has been discovered at the LHC and propose a search strategy to determine its spin based on two simple asymmetries in the $ZZ$, $W^+W^-$ and $t\bar{t}$  decays channels. We consider some benchmark  values for its mass  (in the interval from 182 GeV/$c^2$ to 1 TeV/$c^2$) and discuss the relative advantages of the different  decay processes. A full analysis, including  the background, is given. For a center-of-mass  energy of 14 TeV, we find that the lowest integrated luminosity required to discriminate between the different spins is, depending on the process and the resonance mass, between 40 fb$^{-1}$ and 250 fb$^{-1}$. 

\end{abstract}

\maketitle
\vskip1.5em
\section{Motivations} 
\label{sec:mot}
Our understanding of the breaking of the electroweak (EW) symmetry is closely link to the study of resonances or fundamental states (like the Higgs boson itself) which are produced as a single state in the $s$-channel of proton-proton collisions and then decay into $ZZ$, $W^+W^-$ and $t\bar{t}$ pairs. These processes make the realization of the EW symmetry manifest  because they are sensitive to the interaction vertices that are ultimately responsible for its breaking; their study should therefore reveal   crucial details about the form of these interactions, the knowledge of which is essential for identifying the correct EW Lagrangian and the relevant fundamental degrees of freedom. These may be those of the standard model (SM), with just a fundamental Higgs boson in the Lagrangian, or turn out to be those of a different theory, in particular one without a Higgs boson in the Lagrangian. For this reason the physics of the EW symmetry breaking is best described in terms of an effective Lagrangian  which includes only the fields already known---fermions and gauge bosons---and leaves the effect of the additional states to higher order operators~\cite{eff-th}. While a direct estimate of the coefficients of these higher-order operators is probably beyond the energy reach of the LHC, their values enter in the determination of the resonance masses and widths~\cite{fv} which are expected to lay around the TeV scale.  

More importantly from the experimental point of view, in the energy range explored by the LHC, the effective theory must give rise to one or more states, the effect of which is to restore unitarity in the high-energy  scattering amplitudes. These states may be fundamental---like the Higgs boson in the SM or in a supersymmetric extension of it---or composite, resonances of a strongly interacting underlaying theory. This new theory may be a strongly interacting Higgs sector, or a technicolor extension to the fermion sector, or an even more exotic and yet to be discovered extension to the SM. 

Once the lowest in energy of such resonances has been identified---say, what we may call the ``Higgs boson'' has been discovered---we will know its mass and the next question will be the determination of its properties, in particular spin and other quantum numbers like $C$ and $P$ parities. Only by knowing these quantum numbers it will be possible to relate the data to the possible theoretical models. The most important of these quantum numbers is the spin. Is the resonance a scalar or a vector? Or does it even have a higher spin value? The answer to this question is important in pointing us toward the underlaying theory---be that the SM or another model.

While the SM Higgs boson mass is already restricted by EW precision measurements and direct searches at LEP to be (at the $3\sigma$ level) between 114 and 170 GeV/$c^2$~\cite{PDG}, scalar  resonances, and even fundamental Higgs bosons, in models beyond the SM are not bound to have masses in this range and can be much heavier without violating any known measurement.  There exist several theoretical models in which what we call here the  ``Higgs boson'' is rather heavy either because it is a bound state or because, even though a fundamental degree of freedom, it happens to have a large mass~\cite{heavyHiggs}. 

The constraints on the possible masses of vector resonances depend on their properties. Whereas the masses of those usually described as extra gauge bosons $Z'$ and coupled to all fermions are severely restricted with values that must be,  depending on the specific model, larger than  700 or even 1500 GeV/$c^2$~\cite{PDG}---vector resonances about which we do not assume any specific coupling to the fermions have masses for which these bounds may not apply or are significantly weaker. 

In this work, we do not discuss higher values for the spin.

Accordingly, we study a broad range of possible masses for both the scalar and the vector resonances.  It useful to organize these masses into five possible  ranges (see Table~\ref{masse}) and name  each range according to some ranking; these ranges are conventional and  identified, respectively, by the current LEP limit, the thresholds for the on-shell production of the relevant decay states and the limit in the SM Lagrangian arising from the requirement of  unitarity. 

\begin{center}
\begin{table}[h!]
\begin{tabular}{|c|c|}
\hline
Flyweight        &  $<$ 114 GeV/$c^2$   (above the LEP limit) \\
\hline
Bantamweight &    115-182 GeV/$c^2$  (up to the $ZZ$ threshold) \\
\hline
Featherweight & 182-342 GeV/$c^2$  (up to the $\bar tt$ threshold) \\
\hline
Welterweight& 342-650 GeV/$c^2$  (up to near the unitary limit) \\
\hline
Cruiserweight & $>$ 650 GeV/$c^2$ \\
\hline
\end{tabular}
\caption{Ranking and naming of the mass ranges for the resonances considered in this work.}
\label{masse}
\end{table}
\end{center}

While fly- and bantamweight bosons can be be thought as fundamental, welter- and cruiserweight bosons are best considered as resonances generated as bound states by some strong interaction. The intermediate featherweight bosons can  be both fundamental or  bound states. 

In the following, we explore the whole range above the $ZZ$ threshold, from feather- to cruiserweight bosons, to determine whether the spin is zero or one. Flyweight bosons are essentially excluded by the LEP bound (unless some clever construction is invoked). The spin of bantamweight bosons is difficult to study by our technique because in this region the cross section is computed, by definition, below the threshold for the on-shell production of the decaying states and therefore too small.

Our strategy is to identify some asymmetry---in the cross section for the resonance production and decay into $ZZ$, $W^+W^-$ or $t\bar{t}$ pairs---which is sensitive to the  spin of the resonance and that has different values for the scalar state, the vector state and the background, respectively. These differences make it possible to compare the data and, hopefully, identify the value of the spin by studying a single observable. Asymmetries are nice observables because most of the systematic uncertainties cancel out. Moreover, they are simple both to visualize and be analyzed.

We begin by presenting a Montecarlo simulation for  both the signal and the background for the three decay channels as generated in proton-proton collisions at the LHC with a center-of-mass energy of 14 TeV. The signal consists in the identification of the resonance, that is, in its experimental discovery---which we assume to have occurred. Following such a discovery of the ``Higgs boson,'' we study the asymmetries we have introduced in order to identify its spin. We present a simplified but realistic estimation of the background and its subtraction from the simulated events and give the integrated luminosity necessary to discriminate between the spin-0 and spin-1 cases. These integrated luminosities are the main result of our work. 

A similar analysis in the case of  the $ZZ$ decay channel has been  presented in~\cite{DeRujula, Gao}  by means of a  full angular analysis and multivariate technique. We briefly compare results in the final discussion.  Analyses with some overlap with ours are contained in~\cite{Eboli,Englert,Paver} for the case of di-lepton final state. As far as we know, the $W^+W^-$ semileptonic decay channel is here utilized for the first time to determine the spin of a resonance.

\section{Cross sections and spin asymmetries} 
\label{sec:cross}

We begin by defining the cross sections for the production of, respectively, a spin-0 and a spin-1 resonance which then decay into $ZZ$, $W^+W^-$ or $t\bar{t}$ pairs.
We only write the parton-level cross sections with the understanding that they must then be convoluted by the appropriated parton distribution functions.

We study  the dependence of the cross sections on the angle, which we call $\theta^*$, between the momentum of  one of the initial partons and that of one member of the final pair as computed in the rest frame of the resonance. For the processes with $W^+W^-$ and $ZZ$ pair production, we identify in the  distribution of this angle an asymmetry---which we call the  dartboard asymmetry because it resembles the target board in the game of darts---the study of which can be used in disentangling the  spin of the resonance in the data analysis. In the case of $t\bar{t}$ pair production, it is possible to utilize an even simpler observable, the usual forward-backward asymmetry, defined with respect to the same angle.


\subsection{Spin 0}

The spin-0 state, let us call it $X_0$,  is characterized by a mass $M_0$ and a width $\Gamma_0$. 
We assume that its production mechanism at the LHC is the same as that of the SM Higgs boson, namely through the gluon-gluon fusion process $\mathrm{gg}\to X_0$ via an intermediate top quark loop.
The effective Lagrangian for the $X_0$ boson coupling to gluons is taken to be identical to that of the SM Higgs boson:
\be
\mathcal{L}=-\frac{1}{4} \frac{\alpha_{s}}{3\pi \upsilon}I(x) X_0 G_{\mu\nu}^a G^{\mu\nu}_a\, ,
\ee
where $\alpha_s=g_s^2/4\pi$ and $g_s$ is the strong coupling, $\upsilon=1/(\sqrt{2}G_F)^{1/2}$ is the electroweak VEV, $X_0$ the field of the resonance and $G_{\mu\nu}$ the gluon field strength tensor. $I(x)$ is a dimensionless function of $x=m_t^2/m_{X_0}^2$ given by
\be 
I(x)=3x \Big[ 2+(4x-1)F(x) \Big]\, ,
\ee
where
\be 
F(x)=\frac{1}{2} \, \theta(1-4x)\left[\log\left(\frac{1+\sqrt{1-4x}}{1-\sqrt{1-4x}} \right)-i\pi \right]^2-2\, \theta(4x-1)  \arcsin ^2 \Big[1/(2\sqrt{x}) \Big] \, .
\ee
We also assume that the resonance $X_0$ couples to the gauge bosons $W$ and $Z$  and the top quark in the same way as the SM Higgs boson. The Lagrangian is thus taken to be identical to that of the SM:
\be
\mathcal{L}=g\, m_W X_0W_\mu^+ W^{\mu -} +g\frac{m_Z}{2\cos\theta_W}  X_0 Z_\mu Z^\mu-g\frac{m_t}{2 m_W}X_0 \bar{t}t  \,,
\ee
where $\theta_W$ is the Weinberg angle and $g$ is the $SU(2)$ weak coupling.

Consider first the parton level process $\mathrm{gg}\to X_0\to VV$, where $VV$stands for either $W^+W^-$ or $ZZ$.
We denote by $p_1$ and $p_2$ the momenta of the two incoming gluon and by $k_1$ and $k_2$ the momenta of the two outgoing gauge bosons. In the resonance rest frame (CM) these momenta can be explicitly written as 
\be
\label{momenta}
p_1=(\sqrt{s}/2,0,0,|p|), \quad p_2=(\sqrt{s}/2,0,0,-|p|) , \quad k_1=(E,|k|\sin\theta^*,0,|k|\cos\theta^*) , \quad k_2=(E,-|k|\sin\theta^*,0,-|k|\cos\theta^*)\,,
\ee
where $s=(p_1+p_2)^2=(k_1+k_2)^2$ and $\theta^*$ is the angle between the incoming gluon and the outgoing $W^+$ or $Z$.
The differential cross section can be computed and is given by
\be\label{0ww}
\frac{d\sigma(\mathrm{gg}\to X_0\to VV)}{d\cos\theta^*}\Bigg|_{CM}=\frac{1}{36} \, \frac{1}{(8\pi)^3}\frac{g^2\alpha_s^2m_V^2}{ \upsilon^2a_V}\left|I(x)\right|^2 \frac{1}{(s-m_0^2)^2+m_0^2\Gamma_0^2}\left( 3+\frac{s^2}{4m_V^4}-\frac{s}{m_V^2}\right)\Big[s(s-4m_V^2)\Big]^{1/2}\, ,
\ee 
where $a_V=1$ for $V=W$ and $a_V=2\cos\theta_W$ for $V=Z$.

Let us consider next the parton level process $\mathrm{gg}\to X_0\to t \bar{t}$. As before, the momenta of the two incoming gluons are $p_1$ and $p_2$ and the momenta of the outgoing $t$ and $\bar{t}$ are $k_1$ and $k_2$; as before, their explicit values  are given in the resonance rest frame by \eq{momenta}.
The differential cross section can  be computed and is given in this case by
\be\label{0tt}
\frac{d\sigma(\mathrm{gg}\to X_0\to t \bar{t})}{d\cos\theta^*}\Bigg|_{CM}=\frac{1}{72}\frac{1}{(8\pi)^3}
\frac{g^2 \alpha_S^2 m_t^2}{\upsilon^2 m_W^2}
\left|I(x)\right|^2\frac{1}{(s-m_0^2)^2+m_0^2\Gamma_0^2}
s^{1/2} (s-4m_t^2)^{3/2}\, .
\ee

\subsection{Spin 1}

The spin-1 state, $X_1$, is characterized by a mass $M_1$ and width $\Gamma_1$. We assume that this resonance mimics the extra neutral boson $Z'$ which is present in some extended gauge models. In the so called ``reference model''~\cite{Altarelli} the couplings of the  $Z'$ to quarks and weak gauge bosons are a direct transcription of the corresponding SM couplings.   
Accordingly, its production mechanism at the LHC is through the process of $q\bar{q}$ fusion. The interaction vertex $\bar{q}q X_1$ is assumed to be the same as that of the $\bar{q}qZ$ in the SM, that is
\be
\frac{ig}{2 \cos\theta_W} \gamma^\mu (c_V^q-c_A^q\gamma_5) \, ,
\ee 
where $c_V^q=T^q_3-2Q_q \sin^2\theta_W$ and $c_A^q=T^q_3$.
We also assume that the three-vector boson vertex $X_1VV$ is given by
\be
-ig\, \cos \theta_W[(q-k)_\lambda g_{\mu\nu}+(k-r)_\mu g_{\nu\lambda}+(r-q)_\nu g_{\lambda\mu}]\, ,
\ee 
in the case of $VV=W^+W^-$ and
\be
-ig\,(q_\nu g_{\mu\lambda}+q_\lambda g_{\mu\nu})\, ,
\ee 
in the case of $VV=ZZ$.
In the formulas above all gauge boson momenta are taken to be incoming, the momentum index pair $(q,\mu)$ corresponds to $X_1$, while $(k,\nu)$ and $(r,\lambda)$  to $W^+W^-$ or $ZZ$ respectively.

Consider the parton level process $q \bar{q}\to X_1\to V V$, where $VV = W^+W^-$ or $ZZ$. We denote now by $p_1$ and $p_2$ the momenta of the two incoming  quarks and by $k_1$ and $k_2$ the momenta of the two outgoing gauge bosons. As before, in the resonance rest frame  these momenta are given by \eq{momenta}. The respective differential cross sections are~\cite{Altarelli}
\bea
\label{1ww}
\frac{d\sigma(q \bar{q}\to X_1\to W^+ W^-)}{d \cos\theta^*}\Bigg|_{CM}&=&\frac{g^4}{48\pi}K^q \frac{\left[s(s-4m_W^2)\right]^{1/2}}{(s-M^2_{1})^2+M_{1}^2\Gamma_{1}^2}
\Bigg\{ \frac{s^2}{4 m^4_W}\Bigg[ \left(1-\frac{2m_W^2}{s}\right)^2-\left(1-\frac{4m_W^2}{s} \right)\cos^2\theta^* \nn\\
&-&\frac{4 m_W^4}{s^2} \Bigg] \left[ \frac{1}{4}-\frac{m_W^2}{s}+3\frac{m_W^4}{s^2}\right]+\frac{s}{m_W^2}-4 \Bigg\} \, , 
\eea
 and
\bea
\label{1zz}
\frac{d\sigma(q \bar{q}\to X_1\to Z Z)}{d \cos\theta^*}|_{CM}&=& \frac{1}{128\pi}\frac{g^4K^q}
{4m_W^2} \frac{1}{(s-M^2_{1})^2+M_{1}^2\Gamma_{1}^2} s^{1/2} (s-4m_Z^2)^{3/2}\left(1+\cos^2{\theta^*}\right) \, ,
\eea
where $K^q=(c_V^q)^2+(c_A^q)^2$.

Finally, consider  the parton level process $q \bar{q}\to X_1\to t \bar{t}$ where the momenta of the incoming $q$ and $\bar{q}$ are denoted by $p_1$ and $p_2$ and the momenta of the outgoing $t$ and $\bar{t}$ by $k_1$ and $k_2$. Again, in the resonance rest frame they are given by \eq{momenta}. The differential cross section for the process is
\bea
\label{1tt}
\frac{d\sigma(q \bar{q}\to X_1\to t \bar{t})}{d\cos\theta^*}\Bigg|_{CM}&=&\frac{1}{32\pi }\frac{g^4}{16\cos^4\theta_W}\frac{1}{(s-M^2_{1})^2+M_{1}^2\Gamma_{1}^2}\Big[K^q K^t \left[ s(1+\cos^2\theta^*)-4 m_t^2 \cos^2\theta^*\right] \nn\\
&+& 4 m_t^2 K^q H^t  + 8 D^q D^t \left[s(s-4  m_t^2)\right]^{1/2} \cos\theta^*\Big] s^{-1/2}(s-4m_t^2)^{1/2}\, ,
\eea
where $H^t=(c_V^t)^2-(c_A^t)^2$ and $D^{q/t}=c_V^{q/t}c_A^{q/t}$.

\subsection{The dartboard asymmetry}
\label{subsec:dart}
In the case of the $W^+W^-$ and $ZZ$ decay modes, depending on the spin of the resonance, we have different behaviors of the differential cross section with respect to the angle $\theta^*$ between the momenta of the incoming parton and one of the decaying particles. 

In the case of a spin-0 resonance, \eq{0ww} shows no dependence of the differential cross section on the angle $\theta^*$. 
For a spin-1 resonance, \eq{1ww}, in the limit $s\gg m_W^2$, shows an explicit dependence which we can write as
\be
\frac{d\sigma(gg\to X_1\to WW)}{d \cos\theta^*} = b_1 + c_1 (1-\cos^2\theta^*),
\ee
where $b_1$ and $c_1$ are $\theta^*$-independent coefficients. 

Since the angle $\theta^*$ only enters in the square of the cosine, no asymmetry can be seen in comparing the forward with the backward direction. We therefore    define an alternative   asymmetry ${\cal A}_{DB}^S$ as follows:
\be
{\cal A}_{DB}^{S}=\left[ \int_{-z^*/2}^{z^*/2}\frac{d\sigma}{d\cos\theta^*}d\cos\theta^* - \left(\int_{-z^*}^{-z^*/2}\frac{d\sigma}{d\cos\theta^*}d\cos\theta^* + \int_{z^*/2}^{z^*}\frac{d\sigma}{d\cos\theta^*}d\cos\theta^*\right)\right]\Big/ \left [\int_{-z^*}^{z^*}\frac{d\sigma}{d\cos\theta^*}d\cos\theta^* \right] \, ,
\label{DB}
\ee
where $z^*/2$ is the value of $\cos \theta^*$ defining the border between the different  integration regions.

In the  rest frame of the resonance, this asymmetry represents the difference between the central and the external region in the $\cos \theta^*$ distribution of the decay particles thus resembling the scoring sections of a dartboard.\footnote{A similarly defined asymmetry was first introduced in~\cite{CE-asym} by means of a differently defined angle.} For the spin-0 resonance this asymmetry is zero, ${\cal A}_{DB}^{S=0}=0$, while for the spin-1 resonance (decaying into $WW$) the dartboard asymmetry is positive and given by 
\be
{\cal A}_{DB}^{S=1}=\frac{ 3\,c_1z^{*2}}  {12\,(b_1+c_1)-4\,c_1z^{*2} }  \, .
\ee

\subsection{The forward-backward asymmetry}
\label{subsec:forw}
Also for the $t\bar{t}$ decay mode, depending on the spin of the resonance, we have different behaviors of the differential cross sections with respect to the same angle $\theta^*$ which can be used to define an asymmetry. 

In the case of spin-0 resonance, \eq{0tt} shows no dependence of the differential cross section on the angle $\theta^*$ . For the spin-1 resonance, \eq{1tt} shows the following dependence on the angle $\theta^*$:
\be
\frac{d\sigma(q\bar{q}\to X_1\to t\bar{t})}{d \cos \theta^*} = b_1'+c_1'\cos^2\theta^*+d_1'\cos\theta^* \, .
\ee
where  $b_1'$, $c_1'$ and $d_1'$ are coefficients independent of $\theta^*$.

The linear dependence on $\cos \theta^*$  suggests in this case to utilize the simple forward-backward asymmetry ${\cal A}_{FB}^S$ which can be written as follows:
\be
{\cal A}_{FB}^{S}=\left[ \int_{0}^{z^*}\frac{d\sigma}{d\cos\theta^*}d\cos\theta^* - \int_{-z^*}^{0}\frac{d\sigma}{d\cos\theta^*}d\cos\theta^*\right] \Big/\left[\int_{-z^*}^{z^*}\frac{d\sigma}{d\cos\theta^*}d\cos\theta^* \right] \, .
\label{FB}
\ee
Here $z^*$ is the largest value of $\cos\theta^*$ used to compute the asymmetry.
In the resonance rest frame, this asymmetry represents the difference between the forward and the backward region in the $\cos\theta^*$ distribution of the decay particles. For the spin-0 resonance this asymmetry is again zero, ${\cal A}_{FB}^{S=0}=0$, while for the spin-1 resonance the forward-backward asymmetry is positive and  given by
\be
{\cal A}_{FB}^{S=1}=\frac{3\,d_1' z^*}{6\,b_1'+ 2\,c_1'z^{*2}}\, .
\ee

\section{Event generation and analysis} 
\label{sec:event}

In our analysis we look at  the decays into $ZZ$, $W^+W^-$  and $t \bar{t}$ pairs of the resonances $X_0$ and $X_1$ in the following three decay chains: 
\be
X_{0,1} \to ZZ \to 4l\, , \quad  X_{0,1} \to WW \to l\nu jj \, , \quad \mbox{and} \quad  X_{0,1} \to t\bar{t} \to bl\nu bjj \, .
\ee

The advantage of the semi-leptonic $W^+W^-$ and $t\bar{t}$ channels resides in their higher cross-sections with respect to the purely leptonic decays. This advantage is however offset by the important and non negligible background from those SM processes which mimic the same final signature, that is, events with a one lepton (electron or muon), hadronic jets and missing transverse energy due to neutrinos which escape  detection. In particular the main SM processes to be taken into account are the production of  $t\bar{t}$ and $W^+W^-$ pairs and the associated production of a $W$ boson with an additional number of jets (from 0 to 5) coming from quark or gluon emission. 

In what follows, the QCD background has not been simulated since its contribution should be negligible due to the presence of four leptons or one lepton with relatively high transverse momentum together with a significant quantity of transverse missing energy. 

\subsection{Event generation and detector simulation}

The SM background processes  have been simulated using {\tt MADGRAPH/MADEVENT}~\cite{MADGRAPH}, except for the $W$+jets production, for which {\tt ALPGENv2.13}~\cite{ALPGEN} has been used. {\tt PYTHIAv6.4}~\cite{PYTHIA} has been used to simulate the hadronization of the hard processes produced by the matrix-element generators above. 

The $W$+jets events have been generated separately for jet multiplicity that goes from 2 to 5 and for each leptonic flavor ($e$, $\mu$ and $\tau$). In order to restrict the simulation in the kinematical regions of interest, these background events have been generated applying the following cuts at the generator level: 
\begin{itemize}
\item 20 GeV/$c$ on the lepton transverse momentum, 
\item 20 GeV on the transverse missing energy,
\item 30 GeV/$c$ on the parton transverse momenta in the final state. 
\end{itemize}
Afterwards, the cross-section for $W$+jets has been computed by multiplying the LO value (from the {\tt ALPGEN} output) with a K-factor of 1.15 (in order to rescale them at the NLO) and then multiplied for the MLM-scheme efficiency factor to take into account the parton-jet matching. The result is an inclusive NLO cross-section of 620 pb for each leptonic flavor.
The inclusive NLO cross-sections of the other SM background processes are taken to be 875 pb for the $t\bar{t}$ production~\cite{top}, 114 pb for the $WW$ production~\cite{mcfm} and 18 pb for the $ZZ$ production~\cite{ZZxs}.

Since we want to discuss the problem in a model-independent way, we used the SM Higgs production mechanism (via $\mathrm{gg}$ fusion) and the $Z'$ production mechanism (via $q\bar{q}$ scattering), implemented in {\tt PYTHIAv6.4}, in order to simulate, respectively, the spin-0 and spin-1 resonance production. Only in the case of the spin-1 resonance decaying into a pair of $Z$ bosons, we used an {\em ad hoc} modified version of {\tt MADGRAPH}. 

All the analyses presented hereafter have been done for $pp$ collisions with a center-of-mass energy of 14 TeV, using the {\tt CTEQ6L1} parton distribution functions~\cite{PDF1,PDF2}.

\begin{center}
\begin{table} [h!]
\begin{tabular}{|c|c|}
\hline
Signal  ($M_{X}=250$ GeV/$c^2$) & $\sigma$(pb)$\times$BR  \\
\hline
$X \to WW \to l\nu jj$      &   4.36 \\

$X \to t\bar{t} \to bl\nu bjj$      &   0 \\

$X \to ZZ \to 4l$      &  0.02  \\
\hline
\hline
Signal  ($M_{X}=450$ GeV/$c^2$) & $\sigma$(pb)$\times$BR  \\
\hline
$X \to WW \to l\nu jj$      &   1.91 \\

$X \to t\bar{t} \to bl\nu bjj$      &   0.66 \\

$X \to ZZ \to 4l$      &  0.01  \\
\hline
\hline
Signal ($M_{X}=700$ GeV/$c^2$) & $\sigma$(pb)$\times$BR  \\
\hline
$X \to WW \to l\nu jj$      &   0.28 \\

$X \to t\bar{t} \to bl\nu bjj$      &   0.07 \\

$X \to ZZ \to 4l$      &  $2\times 10^{-3}$  \\
\hline
\hline
Signal  ($M_{X}=1$ TeV/$c^2$) & $\sigma$(pb)$\times$BR  \\
\hline
$X \to WW \to l\nu jj$      &   0.04 \\

$X \to t\bar{t} \to bl\nu bjj$      &   $5\times10^{-3}$ \\

$X \to ZZ \to 4l$      &  $2\times 10^{-4}$  \\
\hline
\hline
SM Background & $\sigma$(pb)$\times$BR  \\
\hline
$t\bar{t} \to bl\nu bjj$    &   390 \\

$WW \to l\nu jj$   &     51 \\

$W \to l\nu$ + jets        &   620 \\

$ZZ \to 4l$    &   0.06 \\
\hline
\end{tabular}
\caption{Cross sections   at 14 TeV center-of-mass energy for the various signals and the main background processes after applying the cuts discussed in the text at the generation level.}
\label{tab:xsec}
\end{table}
\end{center}

Despite the different production mechanisms, the cross sections for the spin-0 and spin-1 cases are taken to be of the same size, namely that of  the $\mathrm{gg}  \to H$ production---as calculated at the NNLO+NNLL+EW with soft gluon re-summation~\cite{Grazzini}. The values of these cross sections are shown in Table~\ref{tab:xsec} for the various channels together with the  cross sections for the most relevant SM background processes. 

Since the total widths of the scalar and vector resonances are in principle quite different, we decided---as a compromise among the various models  usually discussed---to normalize both of them to be  5\% of their mass. 
For the $WW$ and $t\bar{t}$ decays, the experimental resolution on the resonance mass reconstruction is larger than these values, while, in the $ZZ$ channel, where the resolution is better since the leptons are reconstructed with very high precision, the width of the resonance can affect the result, mainly by changing the signal-over-background ratio.

In order to have reliable  shapes for the distributions involved in the analysis, we have generated 5 fb$^{-1}$ for the $W$+jets background, 10 fb$^{-1}$ for the $t\bar{t}$ and the $WW$ backgrounds and about 300 fb$^{-1}$ for the $ZZ \to 4l$  background. At least 100 fb$^{-1}$ have been generated for the various signal samples for the same reason. 

The events thus generated have  been passed into the {\tt PGS4} detector simulator~\cite{PGS}. This tool simulates a generic high-energy-physics collider detector with a tracking system, electromagnetic and hadronic calorimetry, and muon system. The main parameters characterizing the different sub-systems have been chosen to reproduce, within an accuracy of about 10-20\% with respect to a realistic detector simulation,  the performance of the ATLAS detector at the LHC. In particular, the electron and the muon reconstruction is possible for a pseudo-rapidity $|\eta | < 2.5$, the hadronic jet reconstruction for $|\eta | < 5$ and the $b$-tagging identification for $|\eta | < 2.5$ , {\em i.e.}, the limit of the inner tracking system. The  pseudo-rapidity  is here defined as $\eta = -\ln(\tan(\theta /2))$, where $\theta$ is the angle between the final particles and the beam direction in the laboratory frame. The final {\tt PGS} output is given by the various reconstructed physical objects: photons, electrons, muons, hadronic tau decays, and hadronic jets including identification algorithms for jets coming from a bottom quark (the so-called $b$-tagging).

\subsection{Experimental definition of the asymmetries}
\label{subsec:expasymm}

The LHC is a proton-proton collider and  we do not have information on the incoming partons (quarks or gluons). Therefore, the quantities entering the asymmetries described in section~\ref{subsec:dart} and \ref{subsec:forw}  require some redefinitions before we can use them.

After having  reconstructed completely the 4-momentum of the resonance $X_{0}$ or $X_{1}$ in the laboratory frame, we can compute the resonance Lorentz relativistic factors ($\beta_X$ and $\gamma_X$) and  boost all the physical objects used in the resonance reconstruction (leptons, jets, $W$, $Z$ and tops) to the resonance rest frame---which is the reference frame in which the equations of  section~\ref{sec:cross} are defined.  In this reference frame, we can thus compute  the angle $\theta^{*}$---which is the angle between the incident parton (quark or gluon) and the decay products of the resonances ($W^+$, $Z$ and $t$) and the angle that enters in the definition of the asymmetries.

In the case of a symmetric initial state, like the $\mathrm{gg}$ fusion process which generates the scalar resonance, no asymmetry can be seen in any process; this is true for a scalar resonance even if the initial state were asymmetric because the final differential cross section does not depend on the $\theta^*$ angle as shown in eqs.~(\ref{0ww}) and (\ref{0tt}). Instead, an asymmetry can be present for a non symmetric initial state, like the $q\bar{q}$ process used to generate the vector resonance (see \eqs{1ww}{1tt}). 

In the $q\bar{q}$ scattering, it is  impossible to define correctly the quark direction and its 4-momentum. Therefore we decided to assign directly to the partons the 4-momenta of the two beams as computed in the laboratory frame. 

Concerning the identification of the quark with respect to the anti-quark, we decided to look at the values of the pseudo-rapidity $\eta_{X}$ of the reconstructed resonance.  No valence anti-quarks are present in the proton, they have to come from the sea and  have  a lower average momentum along the beam direction with respect to the valence quarks. Therefore the sign of $\eta_{X}$ allows us to determine from which beam the quark comes from. This approximation is true for collisions where the transverse momentum of the resonance is small with respect to the longitudinal momentum, otherwise this choice may dilute the asymmetry. Some of the kinematical cuts used in the analysis take into account this effect and try to minimize it.

The $\cos\theta^*$ variable, used to compute both asymmetries, is evaluated using the scalar product of the 3-momentum vectors of the quark and the $W$ (or $Z$ or top) in the resonance rest frame. To distinguish the electric charge of the $W$ or the top, the charge of the lepton from the $W$ decay is used.

\subsection{$ZZ$ pair production}

The best channel to determine the spin of the resonance in the featherweight and welterweight categories is the decay into a $Z$ boson pairs, where both the $Z$ bosons decay to leptons (electrons or muons). Even though the cross section is small with respect to the $WW$ or $t\bar{t}$ channels (about 3 order of magnitude), the signal is very clear since it involves the presence of four leptons in the final state, and the only relevant SM background is the production of two $Z$ bosons (in the $t$-channel via the exchange of a quark) decaying into four leptons. Instead, in the semi-leptonic $WW$ and $t\bar{t}$ channels, the SM backgrounds described before have a very large cross section preventing the detection of the signal over the background.

We have chosen, as benchmark points for the feather- and welterweight cases, two resonances with a mass of, respectively, 250  and 450 GeV/$c^{2}$. We have identified them through the reconstruction of the two on-shell $Z$ bosons. In particular, our baseline selection consists in  four isolated leptons (4 muons, 4 electrons or 2 muons and 2 electrons) in the final state with $P_{T}>10$ GeV/$c$ and $|\eta| < 2.5$ ({\em i.e.}, the tracking system coverage). We take the invariant masses of the pairs of $e^{+}e^{-}$ (or $\mu^{+}\mu^{-}$) to be  within a window  of $\pm15$ GeV/$c^2$ around the $Z$ mass. The two reconstructed $Z$ bosons must  have  an invariant mass that we take within a window of 25 GeV/$c^2$ (in the featherweight case) or 45 GeV/$c^2$ (in the welterweight case) around the mass of the resonance. The width of this window mass has been chosen to be twice of the simulated resonance width. No further cuts have been applied. The dartboard asymmetry in this case is zero for a scalar resonance, while it assumes a negative value for both the vector resonance and the SM background.

Since we have begun our analysis under the  assumption that the new resonance has already been discovered, we can also safely assume that the background can be estimated with some technique. We do not want to investigate  which method could give the best  estimation---it is an important  topic which is however outside the scope of  the present work. We simply make the assumption that the knowledge of the number of background events and its $\cos \theta ^{*}$ distribution shape is known within an uncertainty which mainly quantifies the systematic uncertainty related to estimation technique. We take this uncertainty to be  5\%. To simulate the effect of this systematic uncertainty on the  asymmetries, we have produced $10^4$ experiments where we have smeared out the background with a Gaussian distribution.  The number of events in each bin of the $\cos \theta ^{*}$ distribution is taken as the mean value of the Gaussian distribution, and the amount of the systematic uncertainty (here 5\%) is taken as the variance. We have then subtracted the distribution thus obtained from the signal plus background distribution. This procedure makes it possible to compute the asymmetry for each experiment with its uncertainty. We quote the mean value (of the $10^4$  experiments) as the final measured asymmetry and the variance as the uncertainty due systematic effects.

\begin{figure}[h]
\begin{center}
\includegraphics[width=2.3in]{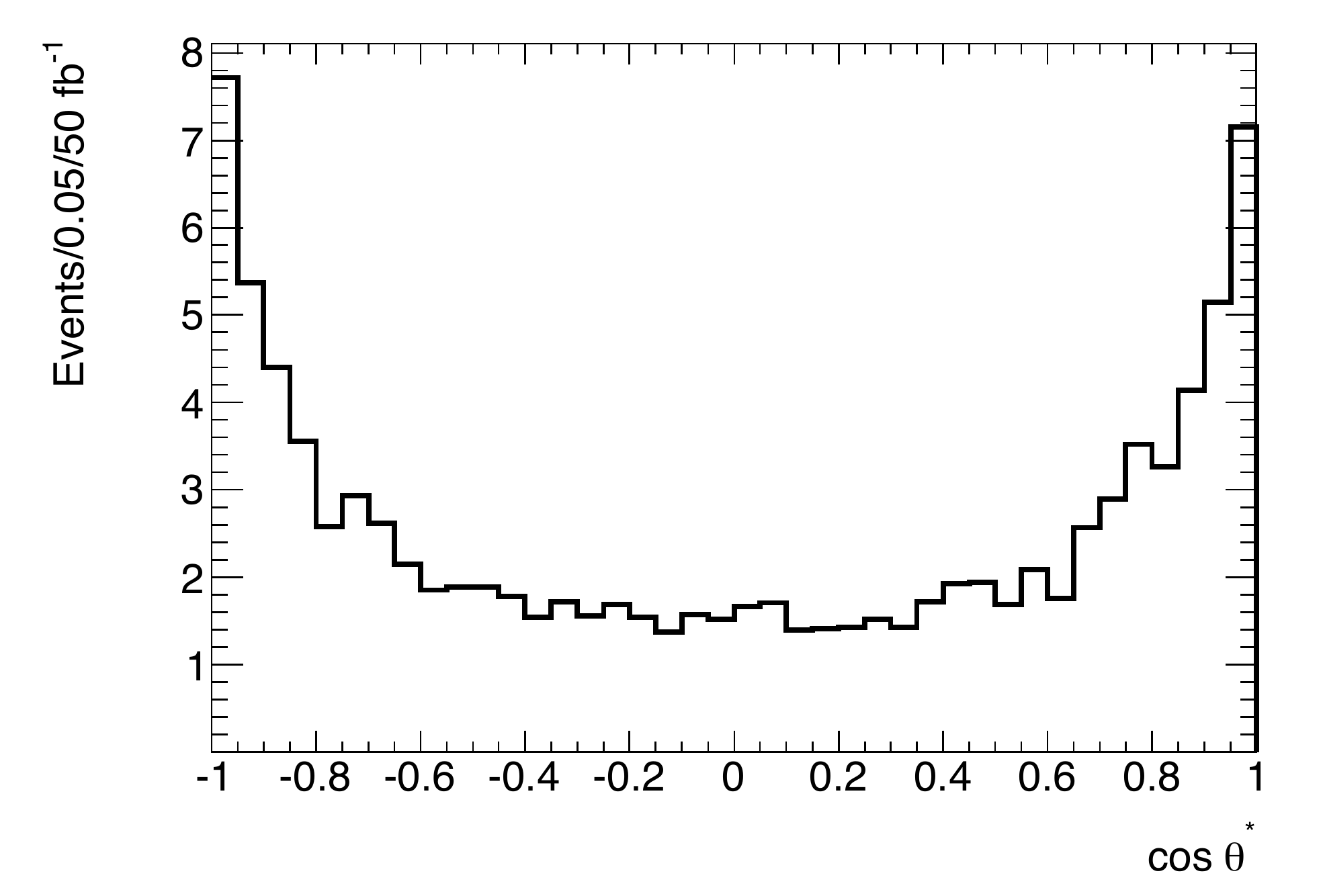}
\includegraphics[width=2.3in]{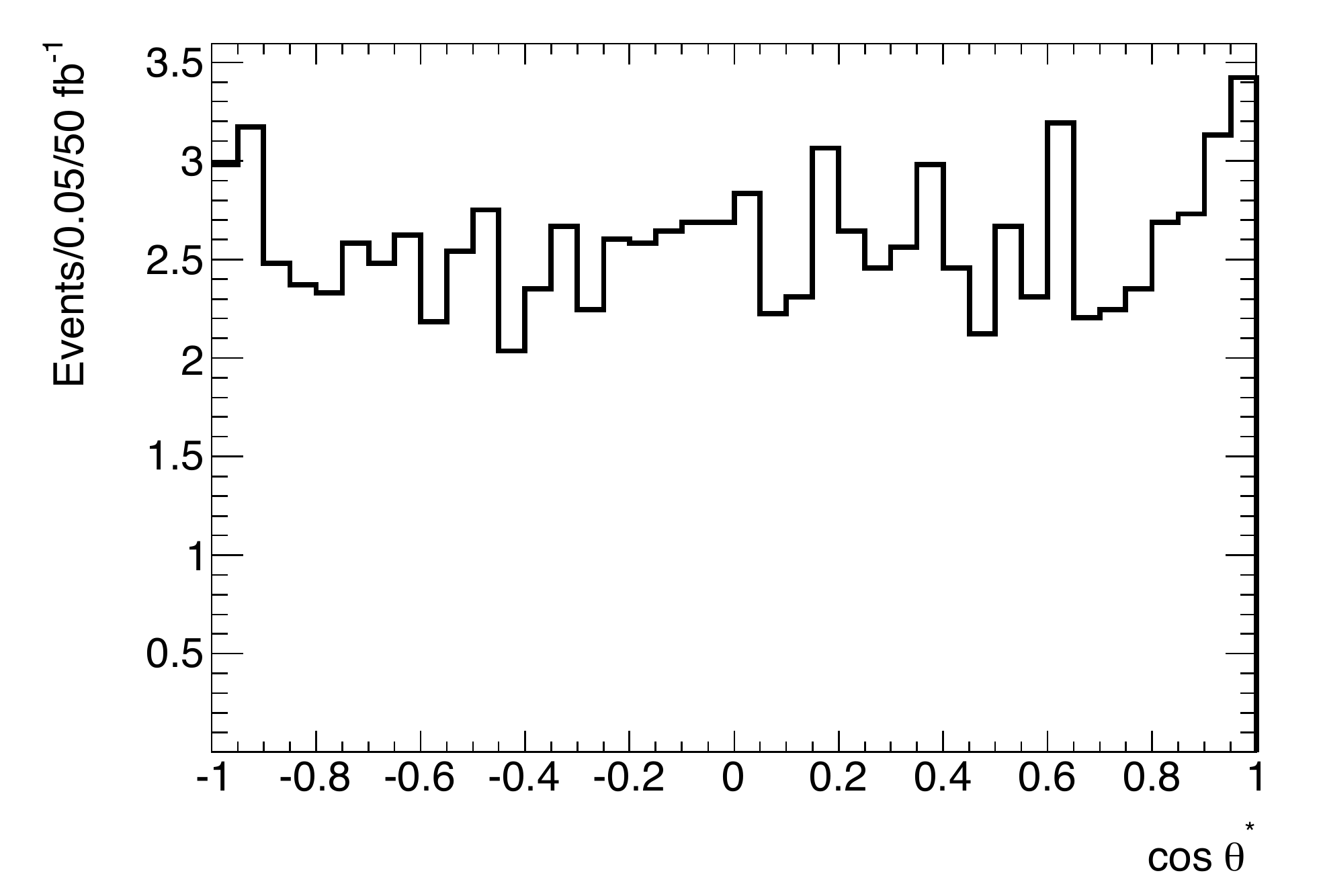}
\includegraphics[width=2.3in]{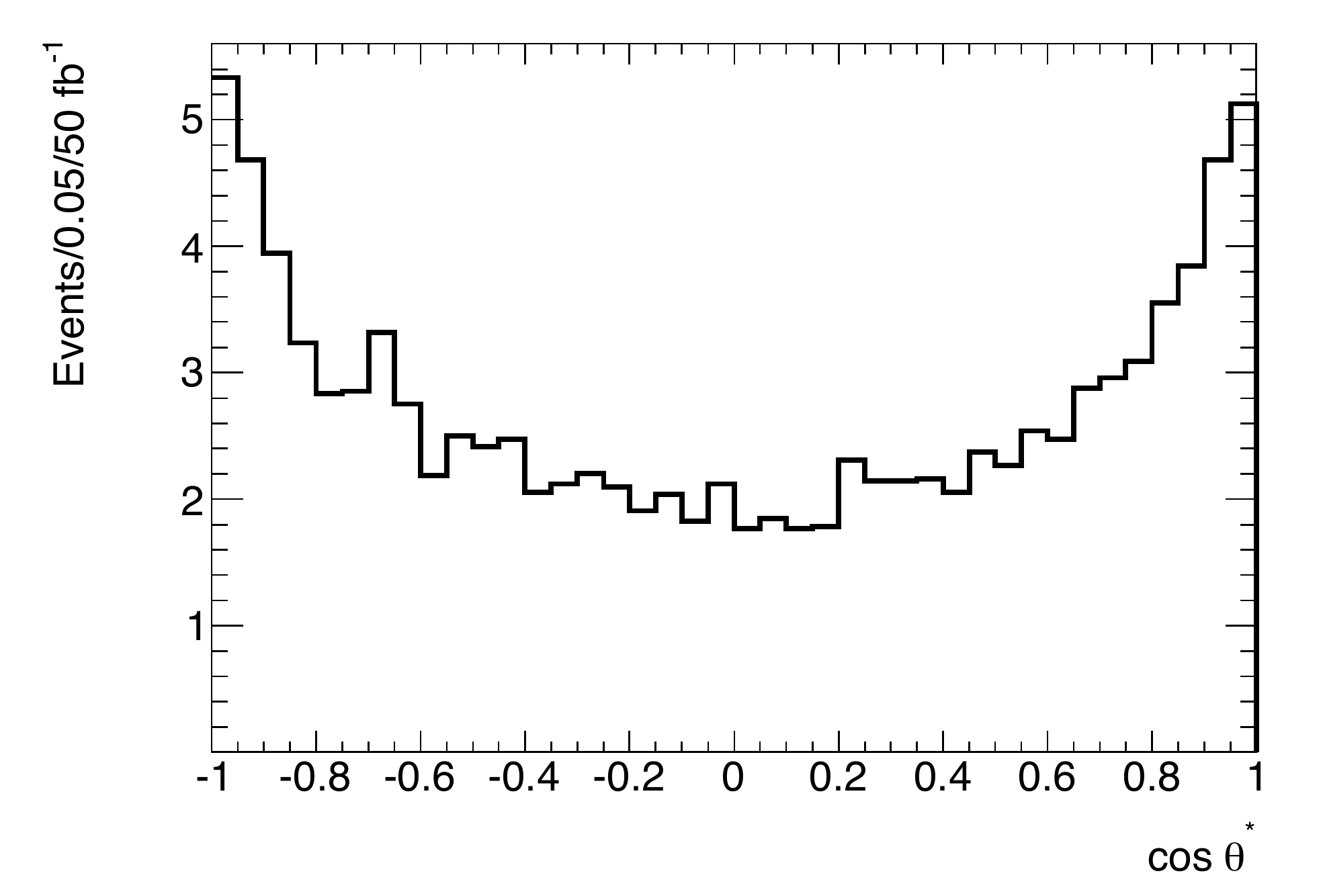}
\caption{Plot of $\cos{\theta^{*}}$ distribution in the $ZZ \to 4l$ channel after the baseline selection (as defined in the text) and the mass window cut for SM (left), scalar resonance (center) and vector resonance (right). The mass of the featherweight resonance is taken to be 250 GeV/$c^{2}$. All the plots are normalized to an integrated luminosity of 50 fb$^{-1}$.}
\label{fig:zz}
\end{center}
\end{figure}

In Fig.~\ref{fig:zz} we show the $\cos \theta ^{*}$ distribution for the signal and the SM background after the baseline selection introduced above and the invariant mass window cut for the  benchmark mass of 250 GeV/$c^2$.

The final number of signal and SM events passing the kinematic cuts, normalized to a statistic of 50 fb$^{-1}$, are shown in Table~\ref{tab:zz1} and Table~\ref{tab:zz2} together with the values of the asymmetry, the statistic and the systematic uncertainties for the two resonance masses analyzed.

For a featherweight resonance (benchmark $m_{X} = 250$ GeV/$c^2$) we have 234 signal events and 203 SM background events. The statistical uncertainty is dominant. With 50 fb$^{-1}$ , we have reached a separation between the scalar and vector cases of $1.7\sigma$, while about 150 fb$^{-1}$ are needed to reach a $3\sigma$ separation in this channel.

For a welterweight resonance (benchmark $m_{X} = 450$ GeV/$c^2$), the signal-over-background ratio is  larger (133 signal events against 32 SM background events), leading to a dartboard asymmetry which is completely dominated by the statistical uncertainty. With an integrated luminosity of 50 fb$^{-1}$, we reach a separation of $1.3\sigma$. In order to separate at the $3\sigma$ level the spin-0 from the spin-1 case, the integrated luminosity must be about 250 fb$^{-1}$.

\begin{center}
\begin{table}[h!]
\begin{tabular}{|c|c|c|c|}
\hline
Signal 50 fb$^{-1}$ & SM events & Signal events &  ${\cal A}_{DB}$ \\
\hline
$X_{0}$ with $m_{X} = 250$ GeV/$c^2$& 203  & 234 & $0.040 \pm 0.065_{\rm stat}  \pm 0.015_{\rm syst}$\\
\hline
$X_{1}$ with $m_{X} = 250$ GeV/$c^2$& 203  & 234 & $-0.112 \pm 0.065_{\rm stat} \pm 0.014_{\rm syst}$\\
\hline
\end{tabular}
\caption{Number of SM and signal events in the $ZZ$ channel passing the final cuts for a resonance mass of 250 GeV/$c^2$. The dartboard asymmetry is quoted with its statistical and  systematic uncertainties. The systematic uncertainty corresponds  to an uncertainty of  5\%  on the SM background.}
\label{tab:zz1}
\end{table}
\end{center}
\begin{center}
\begin{table}[h!]
\begin{tabular}{|c|c|c|c|}
\hline
Signal 50 fb$^{-1}$ & SM events & Signal events&  ${\cal A}_{DB}$ \\
\hline
$X_{0}$ with $m_{X} = 450$ GeV/$c^2$& 32 & 133 &   $0.042  \pm 0.086_{\rm stat}  \pm  0.002_{\rm syst}$ \\
\hline
$X_{1}$ with $m_{X} = 450$ GeV/$c^2$& 32 & 133 &   $ -0.113 \pm 0.086_{\rm stat}    \pm  0.002_{\rm syst}$  \\
\hline
\end{tabular}
\caption{As in Table~\ref{tab:zz1}, for a resonance mass of 450 GeV/$c^2$.}
\label{tab:zz2}
\end{table}
\end{center}

\subsection{$WW$ and $t\bar{t}$ pair production}

For a cruiserweight resonance, the preferred channel  for the dartboard asymmetry separation power is the decay into a $W^+W^-$ pair, where one $W$ boson decays hadronically and the other  in a lepton-neutrino pair (where the lepton is either an $e$ or $\mu$). 

For this mass range, the $ZZ \to 4l$ decay channel, has essentially no SM background, but it is strongly limited by the cross section that is two order of magnitude smaller. This was true also for the feather- and welterweight resonances, as already stressed, but for these mass ranges the kinematic of the signal lies in a corner of the phase space where the SM processes are partially suppressed, thus providing the possibility of identifying the signal from the background with higher efficiency. 

We have chosen as a benchmark two different masses for the resonances: $m_{X}=700$ GeV/$c^2$ and $m_{X}=1$ TeV/$c^2$. 
The baseline selection for the analysis is the request to have in the final state exactly one isolated lepton with transverse momentum $P_{T}>30$ GeV/$c$, the transverse missing energy $E_{T}^{MISS}>30$ GeV and at least two jets with $P_{T}>30$ GeV/$c$ and an invariant mass within a window of $\pm 20$ GeV/$c^{2}$ around the nominal $W$ mass value of 80.41 GeV/$c^{2}$. If more than one pair of jets is inside this window, we keep the nearest one to the nominal $W$ mass value. 

The leptonic side of the event is reconstructed  by assuming that the lepton and the neutrino (and then the missing transverse energy) come from the $W$ decay and forcing the invariant mass of the pair to its mass. This assumption leads to a quadratic equation in the $z$-component of the neutrino momentum ({\em i.e.}, the component along the beam direction). We are  interested only in real solutions and if there are more than one of these, we randomly choose one  in order to avoid bias in the selection. After these cuts, we  can reconstruct both the $W$ 4-momenta and  the invariant mass of the resonance: $m^{2}_{X}=(p_{W_{h}}+p_{W_{l}})^{2}$. 

As before, we assume that the resonance has already been discovered, and we therefore have  at least a rough estimation of its mass. We impose, as a first cut, that the reconstructed invariant mass $m_{X}$ is within a pre-defined window around the measured mass. The width of this window has been taken, as before, to be twice the experimental resolution of the invariant mass peak, in order to be  consistent with the calculations using the "narrow width approximation" and to emphasize the effects of the whole resonance contribution. For both masses, we have chosen a window  of $\pm 120$ GeV/$c^2$ around the reconstructed resonance mass.

\begin{figure}[t]
\begin{center}
\includegraphics[width=2.3in]{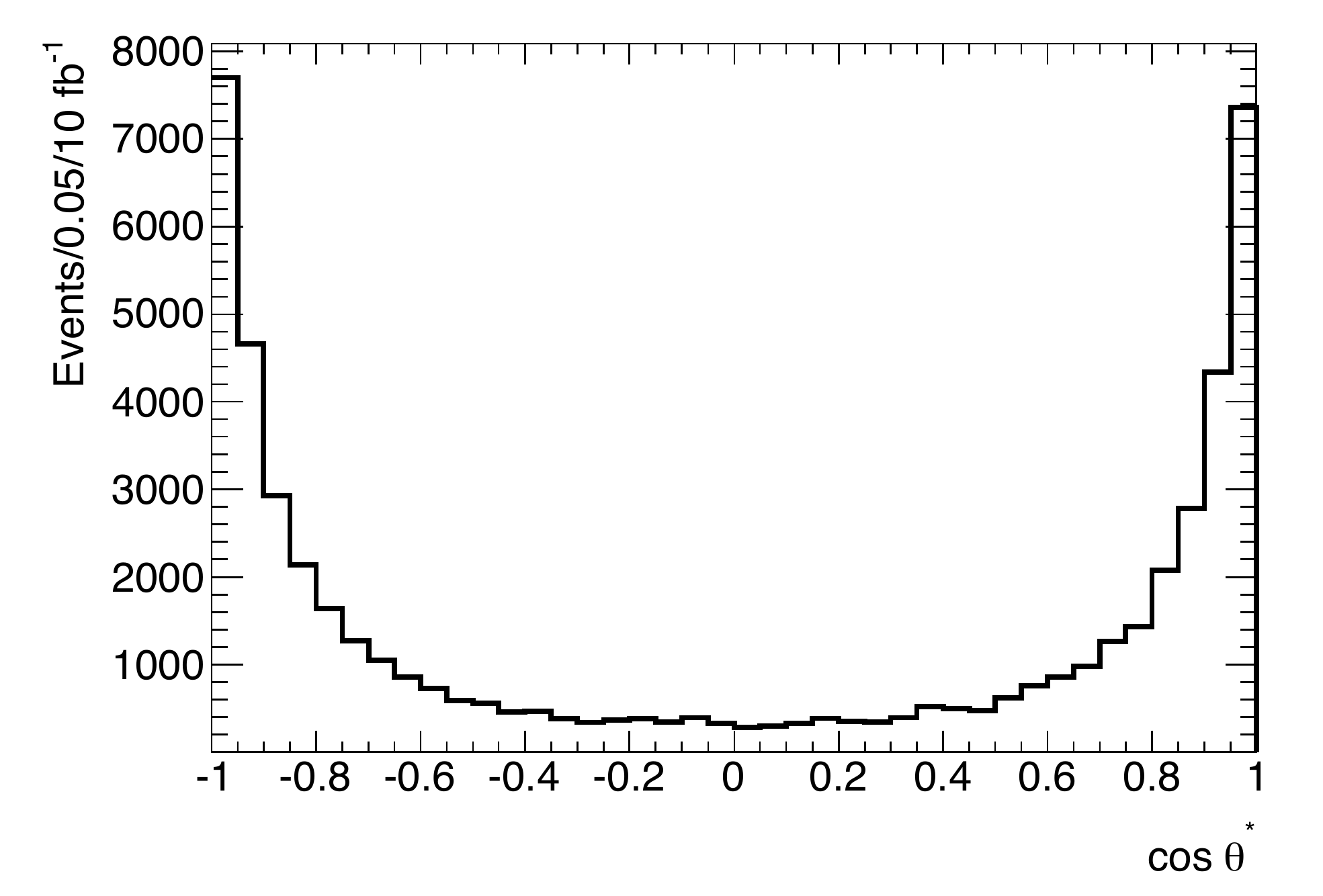}
\includegraphics[width=2.3in]{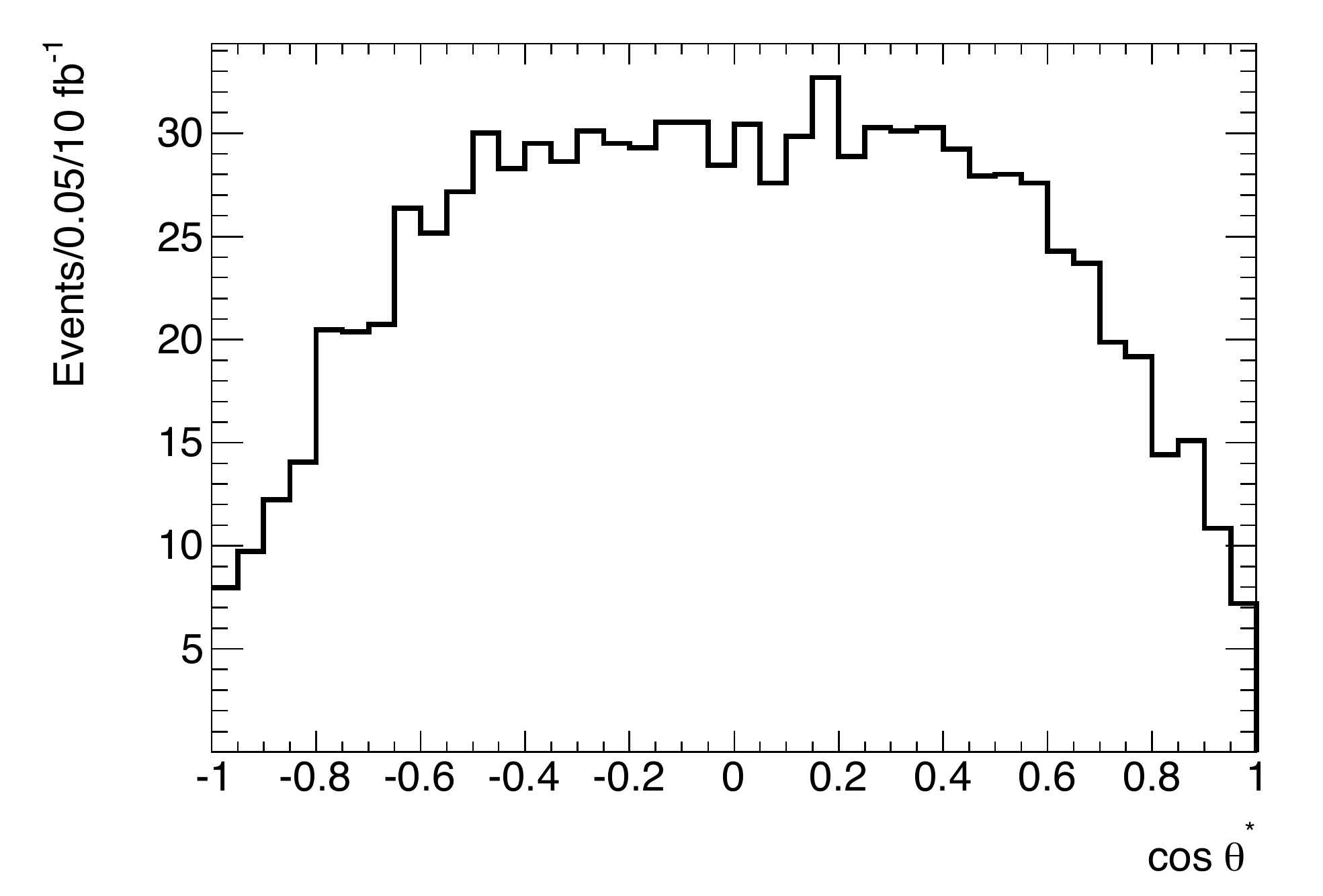}
\includegraphics[width=2.3in]{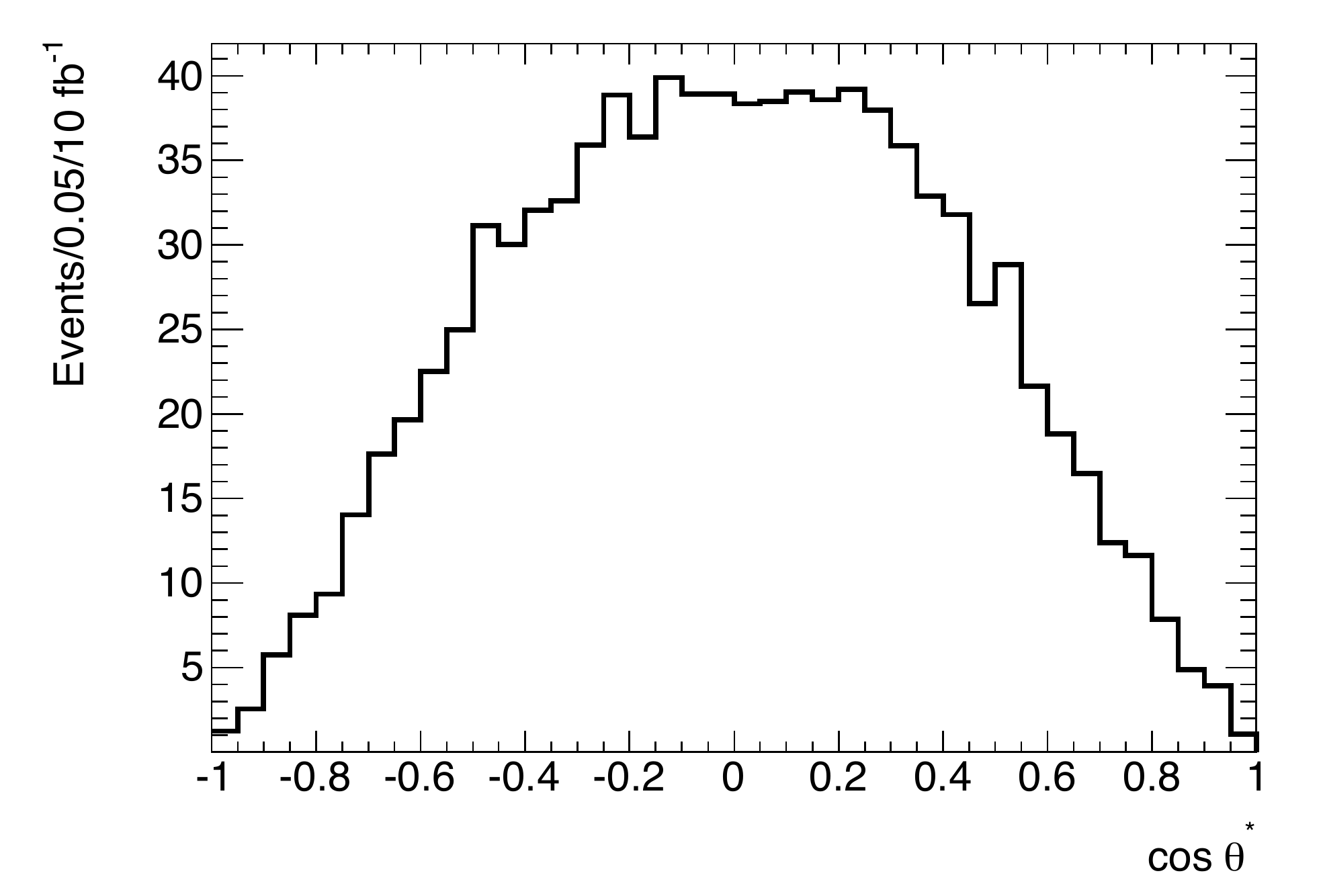}
\caption{Plot of $\cos{\theta^{*}}$ distribution in the $WW \to l\nu jj$ channel after the baseline selection and the mass window cut for SM (left), scalar resonance (center) and vector resonance (right). The mass of the cruiserweight resonance is taken to be 700 GeV/$c^2$. All the plots are normalized to an integrated luminosity of 10 fb$^{-1}$.}
\label{fig:ww}
\end{center}
\end{figure}

In Fig.~\ref{fig:ww} we show the $\cos \theta ^{*}$ distribution for the signal and the overall SM background after the baseline selection and the invariant mass window cut. 

The distribution for the scalar resonance  is not totally flat over the whole range of $\cos \theta ^{*}$, as expected from \eq{0ww}. The smearing for $|\cos \theta ^{*}|>0.7$ is due to the detector acceptance which forbids the reconstruction of the physical objects whose trajectories are very close to the beam pipe. Hence we decided to compute the dartboard asymmetry in \eq{DB} by using $z^*=0.7$.  

Since the selection criteria and the detector acceptance  do affect the angular distribution of $\cos \theta ^{*}$,  we need to found a set of cuts that keeps this distribution as  invariant as possible.
On the other hand, by comparing the cross section of the signal and the main SM backgrounds, one can  notice that the latter are at least three orders of magnitude larger than the former, thus suggesting that a set of hard cuts must be used. What we found---as the best compromise between these two requests---was to use the following four variables to discriminate between the signal and the SM backgrounds:
\begin{itemize}
\item{$H_{T}$=$E^{T}_{MISS}$+$P_{T}^{lept.}$+$\sum P_{T}^{jets}$, where the sum is performed over the jets with $P_{T}>30$ GeV/$c$. Due to the resonance production and decay, the lepton and jets transverse momenta and the transverse missing energy are larger than in the SM case;}
\item{The 3-momentum ${\bf P}_{W}$ of the reconstructed $W$  and the transverse momentum of the reconstructed resonance $P_{T}^{X}$. With respect to the SM events, where no resonances are produced, both the scalar and the vector resonances are quite boosted in the forward and backward regions of the detector, while the $W$ bosons have a larger transverse momentum due to the large mass of their parent state, namely the resonance $X_{0,1}$. Furthermore, the approximation used to determine the quark direction---as explained in section~\ref{subsec:expasymm}---is valid in the regime where the transverse momentum of the resonance is lower than the longitudinal momentum.}
\item{The invariant mass $m_{lj}$ between the lepton and one of the jets in the event. This invariant mass must be larger than a certain threshold $m_{lj}^{thr.}$ A  cut thus defined reduces the $t\bar{t}$ background because, if one jet (not tagged as a $b$-jet) and the lepton come from the same top decay $t\rightarrow bl\nu$, there is a kinematical limit $m_{lj}\leq \sqrt{m^{2}_{t}-m^{2}_{W}}=155$ GeV/$c^{2}$, that is not present if the two objects are uncorrelated (as in the signal events). If two jets are present, we keep the combination with the lowest invariant mass because it is  below the kinematical limit for the $t\bar{t}$ events. The threshold is fixed depending on the mass of the resonance, but it is in any case larger than the kinematical limit above. }
\end{itemize}

Moreover, we impose two additional requests:
\begin{itemize}
\item{No $b$-jets in the event. This is done to suppress  the $t\bar{t}$ background;}
\item{Jet veto. We require at most two hadronic jets with transverse momentum larger than a certain threshold $P_{T}^{veto}$. Since the $W$ bosons coming from the resonance have  high transverse momenta, the jet reconstruction could be problematic due to the granularity of the hadronic calorimeter and the cone algorithm used to reconstruct it. Therefore the hadronic $W$ can be sometime reconstructed as a mono-jet, with a mass (defined as $m_{j} = \sqrt{E_{j}^{2}-|\mathbf{P}_{j}|^{2}}$) that is comparable with the $W$ mass. We  decided to keep these ``$W$ mono-jet events'' if there is a jet with a mass such that $|m_{j}-m_{W}|<20$ GeV/$c^{2}$. In this latter case, the jet veto requires that the $W$ mono-jet is the only jet in the event with $P_{T}>P_{T}^{veto}$. This veto reduces strongly the $W$+jets  and the $t\bar{t}$ background.}
\end{itemize}

 The list of  cuts we applied for the cruiserweight resonance is the following (in parenthesis the cuts for the 1 TeV/$c^2$ case): $H_{T} > 450$ $(650)$ GeV, $P^{veto}_{T} = 30$ $(30)$ GeV/$c$, $P_{W}>200$ $(200)$ GeV/$c$, $P_{T}^{res} < $   50 (40) GeV/$c$ and $m(lj) >$  250 (250) GeV/$c^2$. 

Because of the different kinematics of the scalar and vector resonances, the number of events passing these cuts are different. To make them comparable, we normalize their number to be the same  and then proceed with the computation of the asymmetry. 

The systematic uncertainty on the SM background shape and yield is estimated as in the previous section. This time we quote two possible values for the systematic uncertainty related to estimation technique: a more conservative  value of 10\% and a less conservative of 5\%.

The final number of signal and SM events passing the kinematic cuts, normalized to a statistic of 10 fb$^{-1}$ ({\it i.e.}, one year of LHC running at the nominal instantaneous luminosity $L=2 \cdot 10^{33}cm^{-2}s^{-1}$), are shown in Table~\ref{tab:ww1} and Table~\ref{tab:ww2}  together with the values of the asymmetry, the statistic and the systematic uncertainties for the two benchmark resonance masses.

\begin{center}
\begin{table}[h!]
\begin{tabular}{|c|c|c|c|}
\hline
Signal 10 fb$^{-1}$ & SM events & Signal events & ${\cal A}_{DB}$ \\
\hline
$X_{0}$ with $m_{X} = 700$ GeV/$c^2$ & 677 & 400  &  $0.057 \pm 0.050_{\rm stat} \pm 0.032\,  (0.017) _{\rm syst}$\\
\hline
$X_{1}$ with $m_{X} = 700$ GeV/$c^2$ & 677 & 400  &  $0.215 \pm 0.050_{\rm stat}  \pm 0.032\,  (0.017)_{\rm syst}$ \\
\hline
\end{tabular}
\caption{Number of SM and signal events in the $WW$ channel passing the final cuts for a resonance mass of 700 GeV/$c^2$. The  dartboard asymmetry is quoted with its statistical and  systematic uncertainties. The systematic uncertainty corresponds  to an uncertainty of respectively 10\% and 5\% (in parenthesis) on the SM background.}
\label{tab:ww1}
\end{table}
\end{center}
\begin{center}
\begin{table}[h!]
\begin{tabular}{|c|c|c|c|}
\hline
Signal 10 fb$^{-1}$ & SM events & Signal events & ${\cal A}_{DB}$ \\
\hline
$X_{0}$ with $m_{X} = 1$ TeV/$c^2$& 80 & 73   &  $0.053 \pm 0.120_{\rm stat}  
\pm 0.027\, (0.013)_{\rm syst} $\\
\hline
$X_{1}$ with $m_{X} = 1$ TeV/$c^2$ & 80 & 73   &  $0.191 \pm 0.117_{\rm stat}  \pm 0.029\, (0.014)_{\rm syst} $\\
\hline
\end{tabular}
\caption{As in Table~\ref{tab:ww1} for a resonance mass of  1 TeV/$c^2$.}
\label{tab:ww2}
\end{table}
\end{center}

For a cruiserweight resonance benchmark mass of 700 GeV/$c^{2}$, we found 400 signal events and 677 SM background events. By adding in quadrature the statistical and systematic uncertainties, the possibility to disentangle the scalar  from the vector spin is related to the level of the systematic uncertainty on the SM background. With a conservative evaluation of the systematic level at 10\%, a separation at $2.8\sigma$ ($3.1\sigma$) could be possible with an integrated luminosity of 40 fb$^{-1}$ (100 fb$^{-1}$). On the other hand, an uncertainty of 5\% on the background could give a separation of $2.1\sigma$ already with 10 fb$^{-1}$,  $3.7\sigma$ with 40 fb$^{-1}$ and $5\sigma$ with about 100 fb$^{-1}$, where the systematic uncertainties are dominant.

For a resonance mass of 1 TeV/$c^2$, even if the signal-over-background ratio is better than in the previous case (here we have 80 signal and 73 SM background events),  the rather small cross section requires collecting a higher integrated luminosity  to disentangle the different spin. In fact, with 100 fb$^{-1}$ the statistical and systematic uncertainties become comparable, thus leading to a separation of $2.1\sigma$ ($2.6\sigma$)  for a SM background systematic uncertainty of 10\% (5\%). To reach a $3\sigma$ separation, the estimated integrated luminosity to be collected is about 150 fb$^{-1}$, if the systematic uncertainties reach the 5\% level, and about 300 fb$^{-1}$ in the case of the 10\% accuracy on the systematic uncertainties.

Further information on the nature of the spin of a resonance lying in this mass region can be inferred by analyzing the $X\to t\bar{t} \to l \nu jj$ resonance decay. As explained in section II, in the case of a vector resonance coupled to quarks in a SM-like way, a forward-backward asymmetry ${\cal A}_{FB}$ is present, while it is absent in the scalar and in the SM top pair production processes (see Fig.~\ref{fig:tt} below). The analysis  consists therefore in the extraction and measurement of this asymmetry over a symmetric SM background.

This channel has the same SM backgrounds of the $WW$ semi-leptonic channel, but different composition and mixtures since the $WW$ background is almost negligible after the cuts and the $W$+jets is present for higher jets multiplicity. The cross section is lower with respect to the $WW$ case and also the reconstruction efficiency for top quark pairs is lower than a simple reconstruction of a $W$ boson. There are also some obvious differences in the signature because here we ask for the presence of $b$-jets, 4 hadronic jets in the final state and we have to reconstruct two top quarks instead of two $W$. 

In particular, for the 700 GeV/$c^{2}$ case, we follow the same procedure used to reconstruct  both the hadronic and leptonic $W$ and in addition we ask for exactly two $b$-jets. Then we combine the two $W$ with the two $b$-jets in order to reconstruct the two top quarks by imposing that their masses lie in a window of $\pm$ 40 GeV/$c^2$ around the nominal top quark mass (namely, 172.5 GeV/$c^{2}$). Then the invariant mass of the top quark pair must be in a window of $\pm$ 120 GeV/$c^2$ around the mass resonance value.
For the 1 TeV/$c^{2}$ case, the baseline selection is the same except for the request of having at least one $b$-jet. This choice is motivated by the fact that the $b$-tagging efficiency becomes lower with the increase of the jet momentum and therefore  many signal events are lost. This request lowers also the rejection against the hadronic non $b$-jets, but we can  bypass the problem by hardening some kinematical cuts.

Further kinematical cuts  to be applied lead to the final selection which is the following, respectively, for 700 GeV/$c^2$ and 1 TeV/$c^2$ masses: $H_{T} > $  550 (650) GeV, $P_{top} >$ 200 (350) GeV/$c$ and $P_{j1} >$ 200 (250) GeV/$c$ , where $j1$ is the jet with the highest momentum in the event.

The systematic uncertainties on the SM background have been estimated in the same manner we have previously described, using again 5\% and 10\% as  reference values.

In Fig.~\ref{fig:tt} we show the $\cos \theta ^{*}$ distribution for the signal and the overall SM background after the baseline selection and the invariant mass window cut for the two benchmark masses. In order to limit as much as possible the acceptance detector effects (as in the $WW$ case previously discussed), we decided to compute the forward-backward asymmetry only in the interval defined by $z^* = 0.8$.

\begin{figure}[t]
\begin{center}
\includegraphics[width=2.3in]{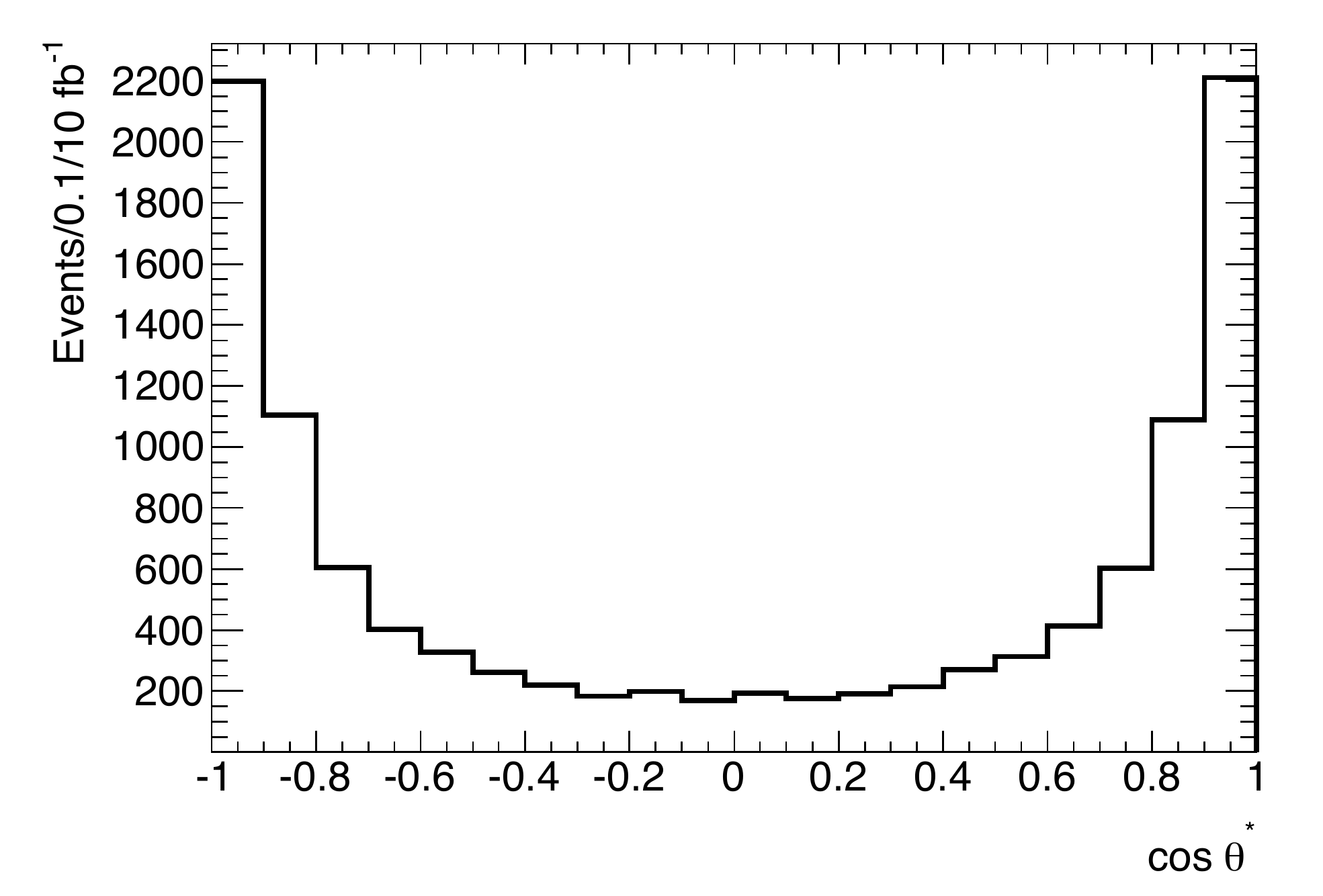}
\includegraphics[width=2.3in]{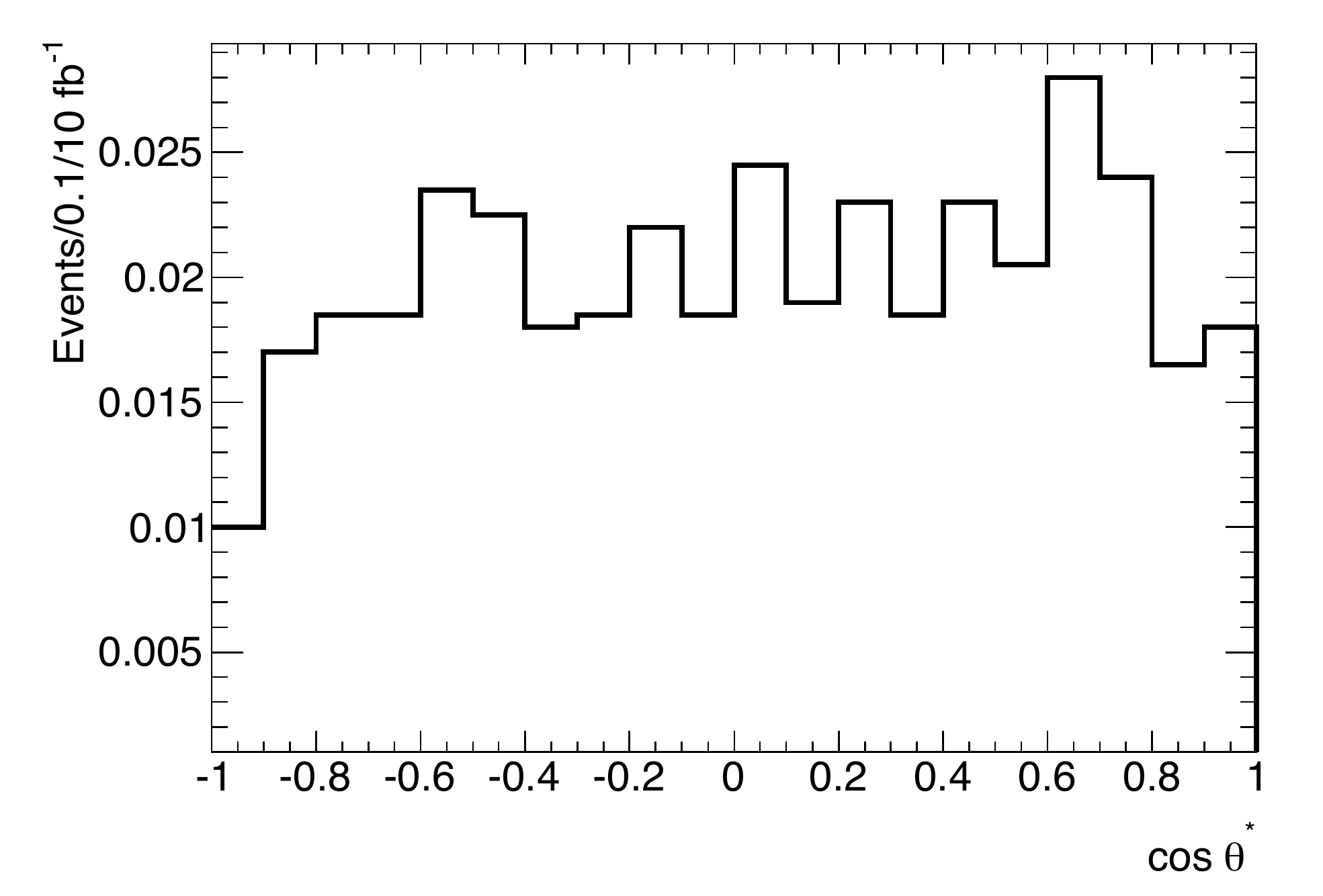}
\includegraphics[width=2.3in]{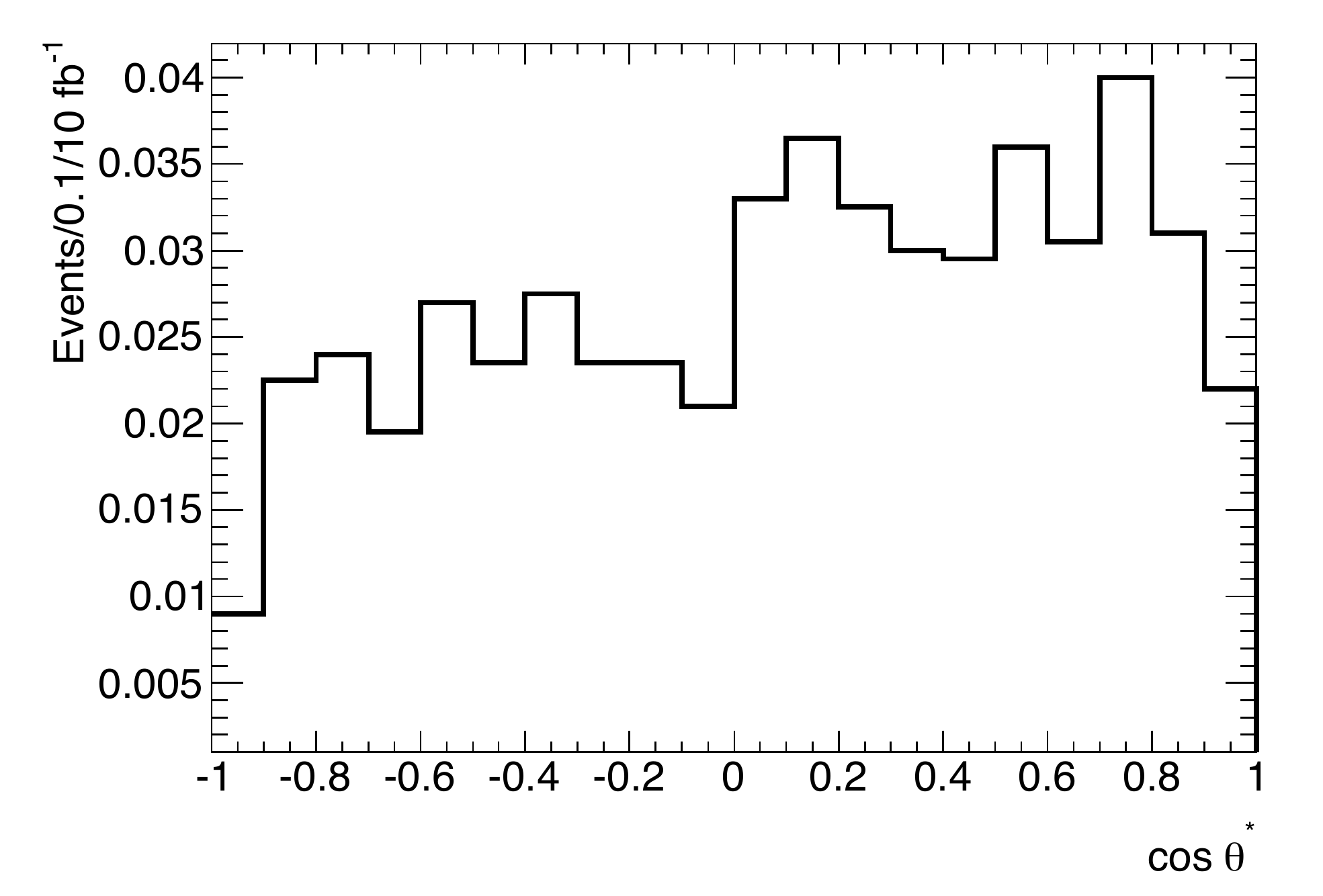}
\caption{Plot of $\cos{\theta^{*}}$ distribution in the $t\bar{t} \to bl\nu$ $bjj$ channel after the baseline selection and the mass window cut for SM (left), scalar resonance (center) and vector resonance (right). The mass of the resonance is 1 TeV/$c^{2}$. All the plots are normalized to an integrated luminosity of 10 fb$^{-1}$.}
\label{fig:tt}
\end{center}
\end{figure}

The non-reducible $t\bar{t}$ background is  dominant, with a cross section about 5000 times larger than the signal. Even though the kinematical cuts above increase by a factor of 10 the signal-over-background ratio, the former is still too small  to find any deviation from the SM. Therefore we decided to make the additional assumption that the $S/\sqrt{B}$ ratio is equal to 5; in other words, a ratio sufficiently large to make possible   the discovery of the resonance in this channel.  This assumption can be considered as conservative, because in presence of systematic uncertainties, this ratio must be larger than 5 to declare a discovery. A rough estimation of the signal cross section (including the branching ratio $X_{0,1} \to t\bar{t} \to l\nu jj$) that would satisfy this assumption, is about 9 pb and 3 pb  for, respectively, a resonance mass of 700 GeV/$c^2$ and 1 TeV/$c^2$.

Under this assumption, the final number of signal and SM events passing the kinematic cuts, are shown in Table~\ref{tab:ttbar1} and  Table~\ref{tab:ttbar2}, together with the values of the forward-backward asymmetry, with the statistic and the systematic uncertainties for the two resonance masses. As a reference, we quote the number of SM events passing our cuts for an integrated luminosity of 10 fb$^{-1}$. The systematic uncertainties are computed again with the procedure explained above, using respectively 10\% and 5\% uncertainties on the SM background.

\begin{center}
\begin{table}[h!]
\begin{tabular}{|c|c|c|c|}
\hline
Signal  $10 fb^{-1}$ & SM events & Signal events & ${\cal A}_{FB}$ \\
\hline
$X_{0}$ with $m_{X} = 700$ GeV/$c^2$& 513 & 114  &  $0.001 \pm 0.095_{\rm stat} 
 \pm 0.118\, (0.058)_{\rm syst} $\\
\hline
$X_{1}$ with $m_{X} = 700$ GeV/$c^2$& 513 & 114 &   $0.178 \pm 0.092_{\rm stat}  \pm 0.118\, (0.058)_{\rm syst} $\\
\hline
\end{tabular}
\caption{Number of SM and signal events in the $t\bar{t}$ channel passing the final cuts for a cruiserweight resonance mass of 700 GeV/$c^2$. The forward-backward asymmetry is quoted with its statistical and  systematic uncertainties. The systematics correspond  to an uncertainty of respectively 10\% and 5\% (in parenthesis) on the SM background.}
\label{tab:ttbar1}
\end{table}
\end{center}
\bigskip
\begin{center}
\begin{table}[h!]
\begin{tabular}{|c|c|c|c|}
\hline
Signal $10 fb^{-1}$ & SM events & Signal events & ${\cal A}_{FB}$  \\
\hline
$X_{0}$ with $m_{X} = 1$ TeV/$c^2$& 558  & 118 &  $0.050 \pm 0.094_{\rm stat} \pm 0.096\, (0.048)_{\rm syst} $\\
\hline
$X_{1}$ with $m_{X} = 1$ TeV/$c^2$& 558  & 118 &  $0.258 \pm 0.090_{\rm stat} \pm 0.097\, (0.048)_{\rm syst}$\\
\hline
\end{tabular}
\caption{As in Table~\ref{tab:ttbar1}, for a cruiserweight resonance mass of 1 TeV/$c^2$. }
\label{tab:ttbar2}
\end{table}
\end{center}

In the assumption that we have already discovered the resonance with an integrated luminosity of 10 fb$^{-1}$,  adding in quadrature the statistical and systematic uncertainties, the possibility to disentangle a non-vanishing value for the  asymmetry  is related to the level of the systematic uncertainty on the SM background. 

With a conservative  value of  10\% for the systematic uncertainty, with an integrated luminosity of 100 fb$^{-1}$ (or even  larger), the systematic uncertainty dominates for both  values of the cruiserweight mass, thus forbidding any attempt to determine a non-vanishing value of the asymmetry.

On the other hand, an uncertainty of 5\% on the background (that is essentially $t\bar{t}$ and therefore  easier, in principle, to estimate with  better precision) could lead  to separable values of the  asymmetry  at the level of about $2\sigma$ with 100 fb$^{-1}$ in the 700 GeV/$c^{2}$ case and around $2.7\sigma$ with 100 fb$^{-1}$  for the 1 TeV/$c^{2}$ case. 
In both cases, the systematic uncertainty on the background remains the major issue.

\section{Discussion} 
\label{sec:concl}
We have studied the production at the LHC of a state  of spin 0 or 1, supposed to represent the Higgs boson or a possible resonance playing a similar role.
Our goal was to determine its spin by means of simple asymmetries in the $ZZ$, $W^+W^-$ and $t\bar{t}$ decay channels, for different values of its mass and taking into account systematic uncertainties on the background events.   We used a dartboard asymmetry for the decays into vector bosons, while  we considered a forward-backward asymmetry for the decays into top quarks. 

To determine the spin of a resonance by means of a single observable like an asymmetry is an attractive idea because it is based on a simple procedure, robust against systematic uncertainties.  

In what we called the featherweight region, we have taken a benchmark resonance  mass of 250 GeV/$c^2$ and  found that the most competitive decay channel is $X \to ZZ\to 4l$. 
The same is true for the welterweight region (benchmark resonance  mass of 450 GeV/$c^2$).
To distinguish the spin-0 case from the spin-1 case---assuming a 5\% systematic uncertainty on the SM background---an integrated luminosity of, respectively, 150 fb$^{-1}$ and 250 fb$^{-1}$ is needed to reach a $3\sigma$ separation between the two cases. 

In the cruiserweight region, we have taken two benchmark  resonance masses of, respectively, 700 GeV/$c^2$ and 1 TeV/$c^2$. Here the most competitive channel is the semi-leptonic decay $X \to WW\to l\nu j j$. 

For a resonance mass of 700 GeV/$c^2$, to distinguish the spin-0  from the spin-1 case---assuming a 10\% systematic uncertainty on the SM background---an integrated luminosity of 40 fb$^{-1}$ is needed to reach a $2.8\sigma$ separation; whereas 
an integrated luminosity of 100  fb$^{-1}$ is needed to reach a $3.1\sigma$ separation. The separation power can be improved if the systematic uncertainty on the SM background is lowered to 5\%; in this case an integrated luminosity of 40 
 fb$^{-1}$ is needed to reach $3.7\sigma$ separation (respectively, of 100  fb$^{-1}$ to reach a $5\sigma$ separation).
For a resonance mass of 1 TeV/$c^2$---assuming a 10\% systematic uncertainty on the SM background---an integrated luminosity of 100  fb$^{-1}$ is needed to reach a $2.1\sigma$ separation; whereas an integrated luminosity of 300 fb$^{-1}$ is needed to reach a $3\sigma$ separation. As before, the separation power can be improved if the systematic uncertainty on the SM background is lowered to 5\%, leading to an integrated luminosity of 100 fb$^{-1}$  to reach $2.6\sigma$ separation (respectively, of 150  fb$^{-1}$ to reach a $3\sigma$ separation).

In the same cruiserweight region, we have also studied the $X\to t\bar{t} \to bl\nu bjj$ decay channel using the forward-backward asymmetry to discriminate between spin-0 and spin-1 resonances. Assuming a conservative SM $H \to t \bar{t}$ cross section for the signal, we found that the SM background completely dominates the event thus making impossible any discrimination. However, in the case where 10 fb$^{-1}$ are sufficient to discover the resonance, we can set some rough lower limit on the signal cross section; for instance, for a cross section of about 3  pb for 700 GeV/$c^2$ (and 9 pb for 1 TeV/$c^2$) resonance mass, and assuming a 5\% background systematic uncertainty, an integrated luminosity of 100 fb$^{-1}$ is needed to reach $2\sigma$ separation between the two spin cases in the first case and $2.7\sigma$ in the latter case. This channel could then be used in combination with that of the $WW$ production in order to improve the separation power of the analysis and  have an indication about the resonance coupling to the quarks.  

How well the simple asymmetries we have introduced fare in comparison with a more comprehensive  analysis based on more than a single angular variable? In~\cite{DeRujula, Gao}, the study of $X\to ZZ \to 4l$ decay channel is performed using a full angular analysis and multivariate technique.  They found a much better result, in terms of separation power, for the feather- and welterweight regions. This is due to the fact that the full information on the final state angles can overcome the low statistics available in this channel, which is the main limitation of the dartboard asymmetry. For the cruiserweight region, instead, the result obtained by means of the dartboard asymmetry  in the $X \to WW \to l\nu j j$ channel  is at least comparable or even better than that based on a full angular analysis in the $X\to ZZ \to 4l$ decay channel---since the former semi-leptonic decay channel has a cross section about two orders of magnitude larger than $ZZ \to 4l$, leaving the dartboard asymmetry separation power essentially limited only by the systematic uncertainties. In this region, the asymmetry we introduced makes it possible to determine the resonance spin with an integrated luminosity  from 40 to 100 fb$^{-1}$ (for the 700 GeV/$c^2$ case) depending on the level of the systematic uncertainty on the SM background estimation.

\acknowledgments

We  thanks C. Duhr, F. Maltoni and O. Mattelaer   for help with {\tt MADGRAPH}.


\end{document}